\documentclass{article}
\usepackage[T1]{fontenc} 
\usepackage[utf8]{inputenc} 
\usepackage{ismir,amsmath,cite,url}
\usepackage{graphicx}
\usepackage{color}

\usepackage{amssymb}
\usepackage[mathscr]{eucal}

\usepackage{graphicx}
\usepackage{subcaption}
\usepackage{tikz}
\usepackage{csquotes}
\usepackage{multirow}
\usepackage{wrapfig}

\newcommand{\drawmatr}[6]{
    \draw[black,fill=#6] (#1,#2,#3) -- ++(#4,0,0) -- ++(0,-#5,0) -- ++(-#4,0,0) -- cycle;
}
\newcommand{\drawtens}[7]{
    \draw[black,fill=#7] (#1,#2,#3) -- ++(#4,0,0) -- ++(0,-#5,0) -- ++(-#4,0,0) -- cycle;
    \draw[black,fill=#7] (#1+#4,#2,#3) -- ++(0,0,-#6) -- ++(0,-#5,0) -- ++(0,0,#6) -- cycle;

    \draw[black,fill=#7] (#1,#2,#3) -- ++(#4,0,0) -- ++(0,0,-#6) -- ++(-#4,0,0) -- cycle;
}
\usepackage{lineno}

\title{Uncovering audio patterns in music with Nonnegative Tucker Decomposition for structural segmentation}

\def\authorname{Axel Marmoret, Jérémy E. Cohen, Nancy Bertin, Frédéric Bimbot}

\multauthor{Axel Marmoret$^1$ \hspace{1cm} Jérémy E. Cohen$^1$ \hspace{1cm} Nancy Bertin$^1$ \hspace{1cm} Frédéric Bimbot$^1$} { $^1$Univ Rennes, Inria, CNRS, IRISA, France.\\
{\tt\small axel.marmoret@irisa.fr}
}

\usepackage[font=small]{caption}
\begin{document}

\maketitle
\begin{abstract}

Recent work has proposed the use of tensor decomposition to model repetitions and to separate tracks in loop-based electronic music. The present work investigates further on the ability of Nonnegative Tucker Decompositon (NTD) to uncover musical patterns and structure in pop songs in their audio form.
Exploiting the fact that NTD tends to express the content of bars as linear combinations of a few patterns, we illustrate the ability of the decomposition to capture and single out repeated motifs in the corresponding compressed space, which can be interpreted from a musical viewpoint. The resulting features also turn out to be efficient for structural segmentation, leading to experimental results on the RWC Pop data set which are potentially challenging state-of-the-art approaches that rely on extensive example-based learning schemes.
\end{abstract}


\section{Introduction}\label{sec:introduction}
A common 
problem in 
Music Information Retrieval domain (MIR) is the design of
musical content representations and features able to capture 
meaningful information in relation to a particular aspect of music. While 
short-term features
are 
dominant in the literature, higher-scale features aiming to describe medium-term patterns and long-term structural properties tend to be much less addressed.

Recent work by Smith and Goto~\cite{smith2018nonnegative} has proposed the use of tensor decomposition to model repetitions in loop-based electronic music, with the purpose of separating tracks in audio content. In this paper, we explore the ability of the method to provide a sparse description of music by capturing and characterizing patterns 
at the bar-scale level in western pop songs in their audio form. As a testbed, we evaluate the effectiveness of the new features for structural segmentation, \textit{i.e.} the task of retrieving the boundaries of the various musical sections (such as verses, choruses, intros, bridges...) which form a music piece.

We first recall (in section 2) the mathematical theory of the tensorial model called Nonnegative Tucker Decomposition (NTD), and we provide detailed illustrations and interpretation of the NTD components on the audio recording of a well-known pop song. We then (in section 3) elaborate on a number of practical considerations related to NTD, which are needed to be taken into account when applying the model to real music data. In the last part of the article (sections 4 and 5), we report on experiments and results obtained with the NTD-derived features for structural segmentation of the RWC Pop Music data set~\cite{goto2002rwc}.






\section{Nonnegative Tucker Decomposition}

\subsection{Time-Frequency-Bar Tensor}
Music in its audio form is often represented in the time-frequency domain as a spectrogram, \textit{i.e.} a 2-dimensional matrix (further denoted as $X$). Along the x-axis, the temporal dimension unfolds, discretized as signal frames, while the y-axis is a frequency-related dimension (such as modules of the Fourier coefficients, pitches, constant-Q transforms, wavelet coefficients...).
\begin{figure}[ht]
\setlength{\belowcaptionskip}{-10pt}%
\setlength{\abovecaptionskip}{-2pt}%
 \centerline{\includegraphics[width=\columnwidth]{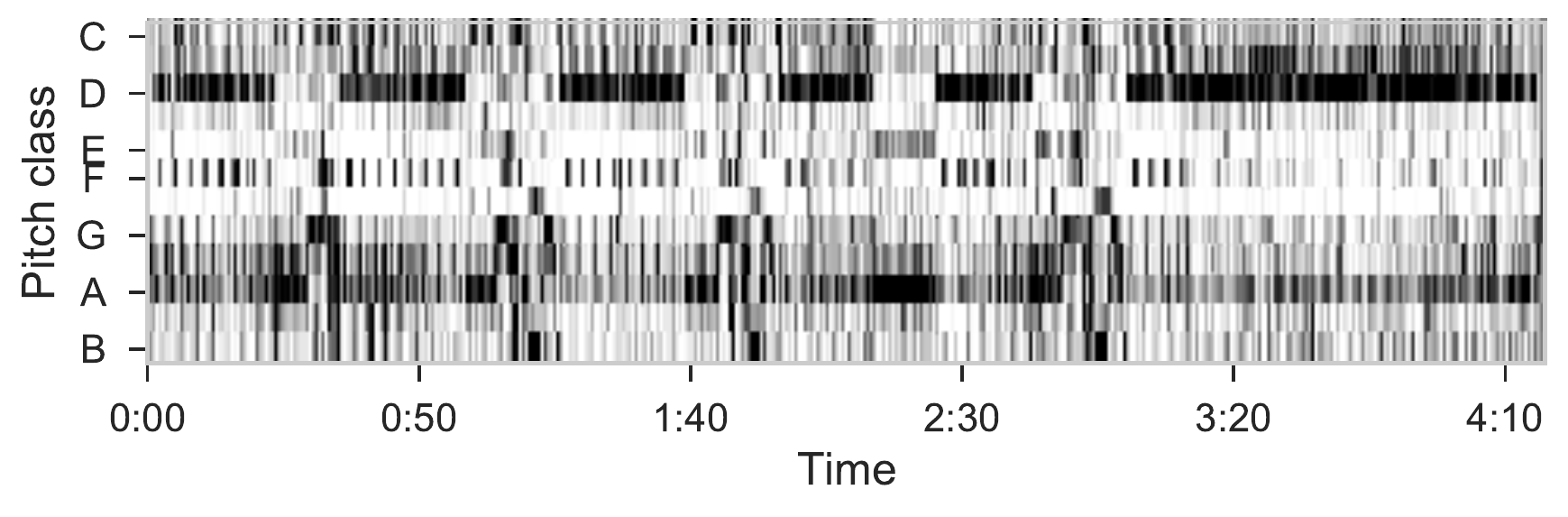}}
 \caption{Chromagram of \enquote{Come Together}, by The Beatles.}
 \label{fig:spectrogram}
\end{figure}

In this work, we describe music as chromagrams. Conventionally, the 12 rows represent the energy distribution across the song for each semi-tone of the classical western music scale, where a note and all its octave counterparts are represented in the same row. An example chromagram is shown in Figure~\ref{fig:spectrogram}.

\begin{figure}[t]
  \begin{subfigure}[ht]{0.3\columnwidth}
 \centering
  \begin{tikzpicture}
    \drawtens{0}{0}{0}{2}{2}{1}{white}
    \draw[->,thick,black] (-0.1,-1.9,0)--(-0.1,-0.1,0);
    \node at (0.7,-0.7,0) {\footnotesize{Pitch-class}};
    \node at (0.5,-1,0) {\footnotesize{Profiles}};
    \draw[->,thick,black] (0.2,-1.9,0)--(1.85,-1.9,0);
    \node at (1,-1.7,0) {\footnotesize{Time at barscale}};
    \draw[->,thick,black] (-0.15,0,-0.3)--(-0.15,0,-1.1);
    \node at (0.7,0.05,-0.6) {\footnotesize{Different}};
    \node at (0.3,-0.04,-0.4) {\footnotesize{bars}};
  \end{tikzpicture}
  \caption{Tensor: multi-dimensional array}
\end{subfigure}
\quad
  \begin{subfigure}[ht]{0.5\columnwidth}
      \includegraphics[width=\columnwidth]{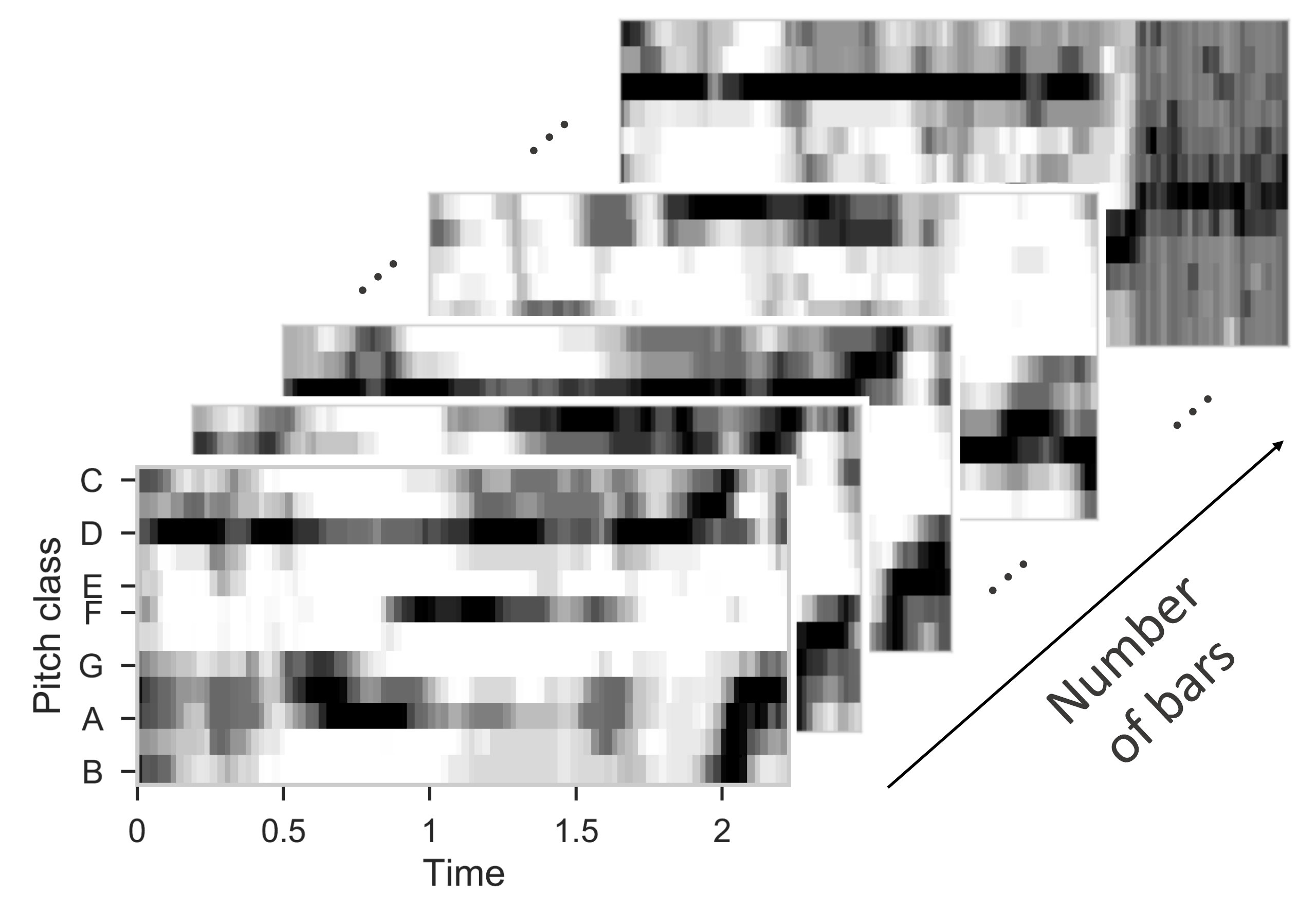}
         \setlength{\abovecaptionskip}{-1pt}%

      \caption{Time-Bar-Frequency representation of \enquote{Come Together} as a concatenation of barwise chromagrams}
      \label{fig:tensor_spectrogram_cap}
  \end{subfigure}
 \setlength{\belowcaptionskip}{-12pt}%
 \caption {Principle and illustration of a Time-Bar-Frequency representation as a third-order tensor.}
 \label{fig:tensor_spectrogram}
\end{figure}
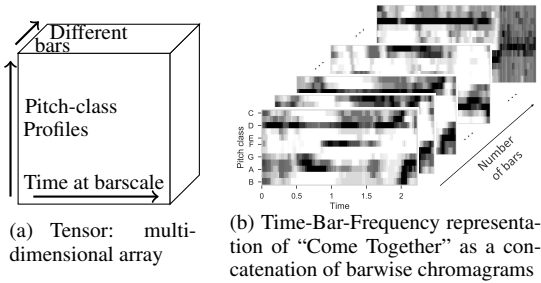

In the tensorial approach, the temporal dimension is broken up into two distinct dimensions: a low-scale dimension representing time in terms of frame index normalized at the bar scale, and a high-scale time dimension representing the bar index within the entire piece. This new viewpoint makes it possible to represent a song as a third-order tensor $\mathscr{X}$ of size $F \times T \times B$
, $F$ being the size of the frequency dimension (12 in the case of chromas), $T$ the number of frames used to describe bars (local time scale) and $B$ the number of bars in the song (global time scale). We call $\mathscr{X}$ the Time-Frequency-Bar representation of the song which, from a data structure viewpoint, is the recasting of $X$ as a 3D array. Tensor $\mathscr{X}$ can be seen as the concatenation of local time-frequency representations, each of them characterizing the content of a bar, as illustrated on Figure~\ref{fig:tensor_spectrogram}. Note that, as bars can be of different lengths in absolute time, the frame hop depends on each bar, and is defined so that all bars contain the same number of frames.

\subsection{Mathematical Model and Formalism}
Let us denote as $\mathscr{X}$  the Time-Frequency-Bar representation of a song, with dimensions $F \times T \times B$. Assuming chroma coefficients are all nonnegative, $\mathscr{X}$ is also a nonnegative tensor. Computing a Nonnegative Tucker Decomposition (NTD) of $\mathscr{X}$ consists in finding 3 nonnegative factor matrices $W$, $H$ and $Q$ (corresponding to the three ``modes'' of the Time-Frequency-Bar tensor) and a nonnegative core tensor $\mathscr{G}$ which relates the three modes as of how to combine them to reconstruct (an approximation of) $\mathscr{X}$.

The dimensions $F' \times T' \times B'$ of the core tensor $\mathscr{G}$ are usually set to be lower than those of  $\mathscr{X}$ (\textit{i.e.} $F',T',B' \leq F,T,B$ respectively). As a consequence, matrices $W$, $H$ and $Q$ are respectively of dimensions $F \times F'$, $T \times T'$ and $B \times B'$ and they can be understood as transformed and compressed representations of the raw information conveyed across the three dimensions of the full tensor.

In conventional tensor-product notation~\cite{kolda_theBible}, the approximation of $\mathscr{X}$ can be written in compact form as:
\begin{equation}\label{NTD}
\mathscr{X} \approx \mathscr{G} \times_1 W \times_2 H \times_3 Q~.
\end{equation}
which rewrites, using element-wise notation, as:
\begin{equation}
\mathscr{X}(f,t,b) \approx \sum_{f',t',b' = 1}^{F', T', B'} \mathscr{G}(f',t',b') W(f,f') H(t,t') Q(b,b')
\end{equation}

\noindent In particular, any given bar of index $b$ 
is represented as:
\begin{equation}
    \mathscr{X}(:,:,b) \approx W\left(\sum_{b'=1}^{B'} Q(b,b') \mathscr{G}(:,:,b')\right)H^T
\end{equation}

\begin{figure}[t]
\begin{tikzpicture}[scale=0.8]
    \drawtens{0}{0}{0}{2}{2}{1.4}{white}
    \node at (2.8,-1) {$=$};
    \node at (1,-1,0){$\mathscr{X}$};
    \draw[->,black] (-0.15,-1.9,0)--(-0.15,-0.1,0);
    \node at (-0.4,-1,0) {\tiny $F$};
    \draw[->,black] (0.1,-2.15,0)--(1.9,-2.15,0);
    \node at (1,-2.35,0) {\tiny $T$};
    \draw[->,black] (-0.1,0.1,-0.1)--(-0.1,0.1,-1.3);
    \node at (-0.2,0.3,-0.4) {\tiny $B$};

    \drawmatr{3.4}{0}{0}{1.2}{2}{white};
    \draw[->,black] (3.3,-1.9,0)--(3.3,-0.1,0);
    \node at (3.1,-1,0) {\tiny $F$};
    \draw[->,black] (3.5,-2.1,0)--(4.5,-2.1,0);
    \node at (4,-2.3,0) {\tiny $F'$};

    \draw[black,fill=white] (5.6,0,0) -- ++(1,0,0) -- ++(0,0,-1.2) --
    ++(-1,0,0) -- cycle;
    \draw[->,black] (5.55,0.1,-0.1)--(5.55,0.1,-0.9);
    \node at (5.45,0.25,-0.4) {\tiny $B'$};
    \draw[->,black] (6.2,0.57,0)--(7,0.57,0);
    \node at (6.6,0.72,0) {\tiny $B$};

    \drawtens{5.2}{-0.5}{0}{1}{1}{0.9}{white}
    \draw[->,black] (5.1,-1.4,0)--(5.1,-0.6,0);
    \node at (4.9,-1,0) {\tiny $F'$};
    \draw[->,black] (5.3,-1.6,0)--(6.1,-1.6,0);
    \node at (5.7,-1.8,0) {\tiny $T'$};
    \draw[->,black] (5.05,-0.5,-0.1)--(5.05,-0.5,-0.9);
    \node at (5.1,-0.2,0) {\tiny $B'$};

    \drawmatr{7.1}{-0.5}{0}{2}{1}{white};
    \draw[->,black] (7,-1.4,0)--(7,-0.6,0);
    \node at (6.8,-1,0) {\tiny $T'$};
    \draw[->,black] (7.2,-1.6,0)--(9,-1.6,0);
    \node at (8.05,-1.8,0) {\tiny $T$};

    \node at (4,-1) {$W$};
    \node at (6.3, 0.2) {$Q^T$};
    \node at (8.05,-1) {$H^T$};
    \node at (5.7,-1) {$\mathscr{G}$};
\end{tikzpicture}
\vspace{-5pt}
\setlength{\belowcaptionskip}{-16pt}%

\caption{Nonnegative Tucker Decomposition of tensor $\mathscr{X}$ in factor matrices $W,H,Q$, and core tensor $\mathscr{G}$, with their dimensions.}
\label{fig:NTD_decomp_tikz}
\end{figure}
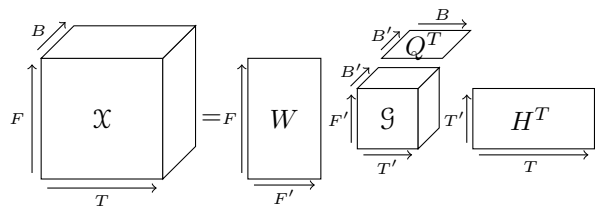

Figure~\ref{fig:NTD_decomp_tikz} depicts a schematic 3-D representation of a NTD. NTD core dimensions $F'$, $T'$ and $B'$ are assumed to be known (or set empirically) prior to the decomposition. As they are lower than the dimensions of their respective mode of the tensor, NTD achieves information compression via nonlinear dimensionality reduction. Indeed, for an original tensor of size $F \times T \times B$ (\textit{i.e.} comprising $F.T.B$ numerical values), the NTD decomposition will total $F.F' + T.T' + B.B' + F' .T'.B'$ values. For example, in the decomposition presented further in Figure~\ref{fig:NTD_figures}, the original tensor contains 102528 real positive values ($F,T,B = 12,96,89$), while only 3626 for the NTD ($F',T',B' = 12,12,10$).

\subsection{Interpretation of the NTD}\label{sec:NTD_interpretation}

Loosely speaking, music can be viewed as composed of musical events (notes, percussive sounds, ...) occurring non-randomly in bars. Under that assumption, bars can be modeled as the combination of a limited set of time-frequency templates along time within a bar, according to some 
rhythmic values, such as 
\enquote{half notes} or \enquote{beamed eight notes} for example. This is the purpose of a conventional musical score, where the major part of symbolic information represents pitch and rhythm. Following that idea, it is a very popular goal in MIR to design methods for turning back musical content (in audio form) into a sparse combination of musical events and temporal activations, as is the case, for instance, with Nonnegative Matrix Factorization (NMF) for music transcription~\cite{smaragdis2003non}. Moreover, music often contains repetitions: different bars can entirely or partly share similar content. For instance, beside almost identical repetitions, some instrumental lines can reoccur in different contexts: an identical bass line in a verse and in a guitar solo, for example.

Combining these observations, we assume that each bar can be represented as the nonnegative combination of a few ``musical patterns'' (as NMF would do), where a \enquote{musical pattern} is itself a sparse combination of musical events and rhythmic activations at the bar scale (for example a melodic line, or a drum fill). Repetitions imply that some of these musical patterns should appear in several bars across a piece. NTD offers an ideal framework to model these properties for music decomposition, musical patterns being efficiently and sparsely shared across bars.

\begin{figure*}[ht]
  \centering
    \begin{subfigure}[ht]{1.95\columnwidth}
        \includegraphics[width=\columnwidth]{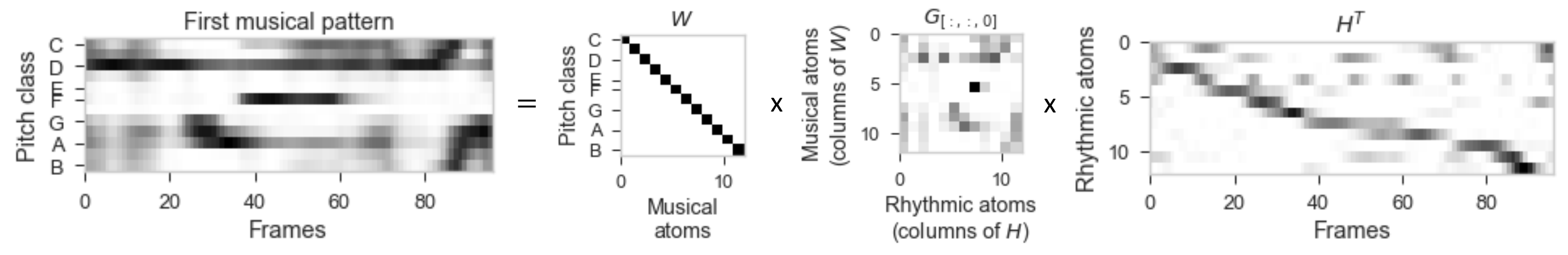}
        \caption{Visual representation of the dominant musical pattern (far-left) corresponding to the first bar of \enquote{Come Together} [a mix of the bass and guitar lines of the intro of the song]. It is decomposed as a linear combination of the columns of $W$ (center-left) representing the chroma information, and of $H$, factors of the rhythmic information (far-right). The musical pattern is itself a linear combination of columns of these matrices, the weights of which are given by the corresponding slice of $\mathscr{G}$ (center-right).}
        \label{fig:musical_pattern}
    \end{subfigure}
    \begin{subfigure}[ht]{1.95\columnwidth}
        \includegraphics[width=\textwidth]{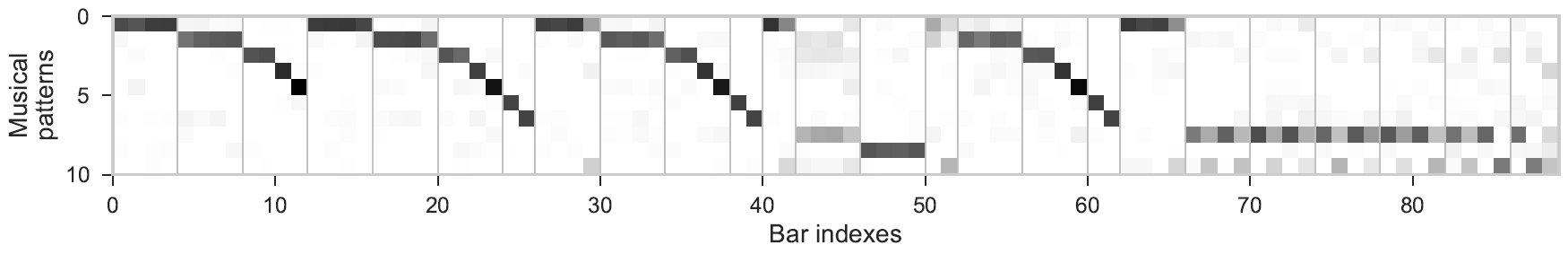}
         \setlength{\abovecaptionskip}{-1pt}%

         \setlength{\belowcaptionskip}{-6pt}%
        \caption{``Come Together'' represented as its $Q^T$ matrix. Each row represents a musical pattern, in their order of appearance in the piece. Grey lines represent the segmentation annotation.}
        \label{fig:Q_mat}
    \end{subfigure}
    \setlength{\belowcaptionskip}{-12pt}%
    \caption{Visualizations of the NTD of Come Together by The Beatles, with ranks $T' = 12$ and $B' = 10$.}\label{fig:NTD_figures}

\end{figure*}

In the NTD, the $W$ matrix represents the musical events, such as the most recurrent notes or chords. 
$H$ represents rhythmic activations at the bar scale, for example 4 quarter notes on the beats. Then, each 2D ``slice'' of the core $\mathscr{G}$ linking these two matrices defines a musical pattern, as a linear combination of some of their columns (musical and rhythmic atoms): for example bass drum hits on the on-beats and snare hits on the off-beats. Finally, $Q$  indicates, for each bar, the combination of musical patterns forming it (generally a few) and their respective intensity.

Figure~\ref{fig:NTD_figures} provides a detailed example of the various NTD components stemming from ``Come Together'' by the Beatles. While the upper part of the figure illustrates the dominant musical pattern for the first bar together with its decomposition in NTD, the lower part depicts the description of the entire piece via the $Q^T$ matrix of the NTD. This example has been obtained from the chromagram represented in Figure~\ref{fig:spectrogram}. Because the song is expressed on the 12-chroma scale, we expect little compressibility with respect to this dimension. We hence simplify the model by fixing $W$ to the 12-size identity matrix. This means that each semi-tone is represented by one and only one column of $W$. For higher dimensions or different representations, columns of $W$ could represent a wider range of harmonic or percussive sounds, chords, or any other frequency pattern. Conversely, ranks $T'$ and $B'$ (respectively the second dimension for $H$ and $Q$) are adjustable parameters of the model. In the decomposition presented in Figure~\ref{fig:NTD_figures}, they have been set to $T' = 12$ and $B' = 10$. All columns of $H$ and all slices of the core linking 
the factor matrices $W$ and $H$, which define the musical patterns, are $l_2$ normalized (\textit{i.e.} divided by their standard deviation).

\section{Practical insights on the NTD}

In this section, we discuss a number of considerations which are bound to have an impact on the actual result of the NTD-based representations and must therefore be taken into account in practical situations.

\subsection{NTD Algorithm}\label{sec:NTD_algorithm}

NTD can typically be computed by minimizing the following non-convex objective function with respect to the nonnegative matrices $W, H, Q$ and the core tensor $\mathscr{G}$:
\begin{equation}\label{NTD_objective_function}
\|\mathscr{X}-\mathscr{G} \times_1 W \times_2 H \times_3 Q\|_F^2
\end{equation}

While a direct global minimization of~\eqnref{NTD_objective_function} is not tractable in general, a standard approach in the literature is to resort to alternating optimization. Following~\cite{phan2011extended}, we solve \eqnref{NTD_objective_function} for $W$, $H$, $Q$ and $\mathscr{G}$ alternatively. It can be shown that each of these steps means solving a matrix nonnegative least-square problem of the form:
\begin{equation}
    \min_{Z\geq 0} \|Y - AZ \|_F^2
\end{equation}
for some matrices $Y, A, Z$. This problem is convex, and it is possible to solve it exactly, or up to an arbitrary precision.
An efficient algorithm for solving matrix nonnegative least squares with high precision is the Hierarchical Alternating Least Squares, and we used an accelerated variant of it to speed-up computation~\cite{gillis2012accelerated}.
The problem of updating $\mathscr{G}$ is also a nonnegative least squares problem, but not a matrix one. Therefore, to update the core tensor $\mathscr{G}$, we used a proximal gradient with optimal step~\cite[Ch.~10]{beck2017first}.

It can be shown that the proposed alternating algorithm is guaranteed to converge to a stationary point of the objective function~\eqref{NTD_objective_function}, since it boils down to an alternating proximal gradient algorithm with optimal step~\cite{bolte2014proximal}. In practice, we used a stopping criterion based either on a fixed maximal number of iterations or on a fixed minimal tolerance of improvement between two successive updates. The entire code, along with experimental notebooks, are published and open-source\footnote{https://gitlab.inria.fr/amarmore/musicntd/-/tree/0.1.0}.
Under this implementation, computing the NTD for ``Come Together'' (4'16'' song) with our algorithm takes approximately 15 seconds on a laptop with an Intel\textsuperscript{\textregistered} Core(TM) i7 processor and 16GB of RAM.

\subsection{Robustness of the NTD}
At least two issues with the NTD make the output of any algorithm highly dependent on the initialization. First, there might be several solutions $W,H,Q,\mathscr{G}$ that provide the same (or a very similar) estimate $\hat{\mathscr{X}}\approx \mathscr{X}$. This problem, known as identifiability deficiency, has been little studied for NTD, and established identifiability conditions are very restrictive~\cite{zhou2015efficient}. Moreover,
these conditions are hard to check in practice. Therefore it is unreasonable to assess the identifiability of the NTD in our application. As a consequence, this means that there might be infinitely many solutions to minimizing \eqnref{NTD_objective_function} that are, from an optimization point of view, equally satisfying. 
Second, even in the case where the NTD is identifiable, the cost function of~(\ref{NTD_objective_function}) is highly non-convex, and local algorithms can only hope to recover a local minimum at best.

These two issues combined give rise to a high dependency of the solution on the initial condition:  from two different initializations, two different results -- most probably non-identifiable local minima -- are likely to be obtained. We have observed such situations in our investigations, with various initializations indeed resulting in different outputs. However, in most cases the decomposition would provide results that were reasonably interpretable from a musical perspective. In particular, when we initialized the algorithm with the absolute values of the Higher Order SVD~\cite{de2000multilinear} computed with the Tensorly toolbox~\cite{kossaifi2019tensorly}, the procedure consistently provided satisfying results for segmentation, as detailed further in our experiments. 

\subsection{Rank Selection}\label{sec:rank_selection}
The ranks $F'$, $T'$ and $B'$ of the decomposition are crucial parameters of the NTD model. Indeed, low ranks tend to over-compress information in the data, failing to uncover relevant structural information in the song, while high ranks may give too much importance to details in the data, resulting in the unability of the model to group similar patterns in a same class of representations.

As developed further in section~\ref{sec:final_results}, our experiments indicate that the optimal ranks are probably specific for each song, which can be easily understood as a consequence of the diversity of intrinsic variability across music pieces. Providing an efficient 
method for selecting the ranks is a challenging topic, left to future work.

\section{NTD-based segmentation}
To study further the relevance of the NTD representation, we evaluated it in the context of structural segmentation. To our knowledge, this is the first attempt to exploit tensorial representations for this purpose.

\subsection{Autosimilarity for Describing Structure}

The autosimilarity matrix $X^T X$ of a music piece ($X$ being its time-frequency representation) is commonly used in structural segmentation. Indeed, similar portions of the piece are likely to have high correlation values. A high density of high values around the diagonal is expected in passages with strong internal similarities, whereas low local correlations would indicate a change in homogeneity. In the ideal case, structural segments appear as consistent blocks with a high level of internal correlation while segment boundaries are points connecting such blocks, surrounded by zones of low cross-correlation.

Nonetheless, music signals usually generate dense autosimilarity matrices, as dissimilar segments in the musicological/perceptive sense (for example a guitar line on the chorus opposed to one in the verse) may still be close in terms of signal properties. While similar parts generate high correlation blocks, it can be harder to characterize segments boundaries when the same instruments are played in all segments (even when playing different lines).


In the present work, we replace $X^TX$ by an autosimilarity matrix $\tilde{Q}\tilde{Q}^T$ computed from the row-wise normalized $Q$ matrix (denoted $\tilde{Q}$), and study its capacity to provide an efficient representation for structural segmentation.

Our assumption is that bar descriptions provided by $\tilde{Q}$ provide a better contrast between similar and dissimilar musical constituents. For instance, we expect two different lines of the same instrument to generate different musical patterns, resulting in lower similarity, whereas compressive effects of NTD will increase correlation of similar events in the transformed space. In that sense, NTD can be seen as a way to uncover piece-dependent features for describing bars, which can then be used to group the bars according to their relative similarity.

\begin{figure*}[ht]
 \centerline{\includegraphics[width=1.9\columnwidth]{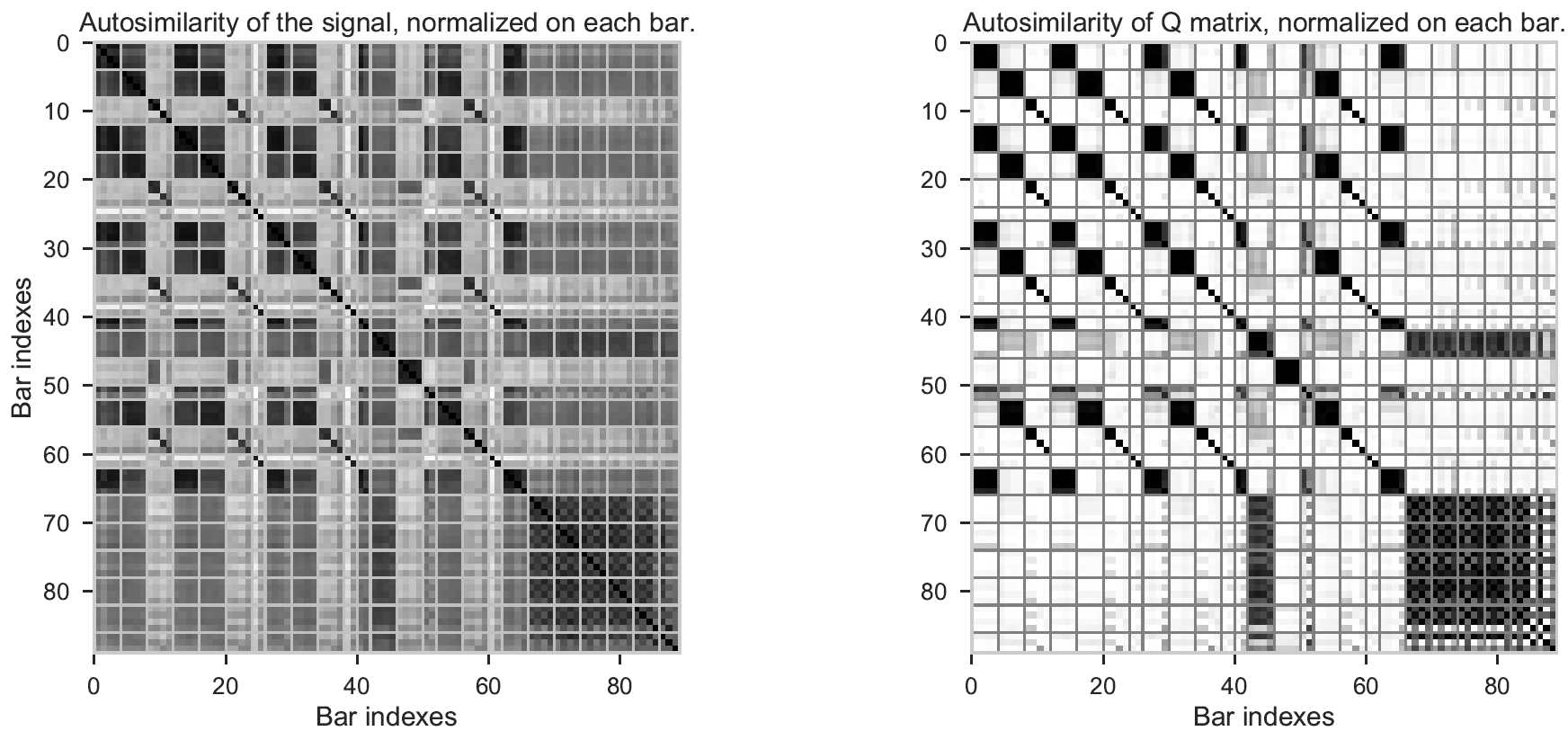}}
 \setlength{\belowcaptionskip}{-12pt}%
 \caption{Barwise $l_1$-normalized autosimilarity matrices for \enquote{Come Together} (0: white - 1: black). Left: barwise chromagram autosimilarity - Right: autosimilarity of $\tilde{Q}$ matrix from Figure~\ref{fig:Q_mat}. 
 Grey horizontal and vertical lines represent the segmentation annotation.}
 \label{fig:autosimilarities}
\end{figure*}

Figure~\ref{fig:autosimilarities} depicts the ``barwise'' autosimilarity matrix of the chromagram $X$ of ``Come Together'': the content of each bar of the signal has been vectorized, and similarity is computed between these barscale vectors. This matrix is compared to the autosimilarity of the $\tilde{Q}$ matrix, presented on Figure~\ref{fig:Q_mat}. This figure visually supports the hypothesis that autosimilarity matrices are sparser when computed from the matrix $Q$ rather than from the chromas $X$. Still, 
highly similar blocks seem to be preserved.

\subsection{A Segmentation Algorithm Using Autosimilarity}
\label{sec:segmenting_autosimilarity}
To assess this hypothesis, we implemented a segmentation algorithm based on the principle of a sliding convolution kernel along the diagonal of the autosimilarity matrix.

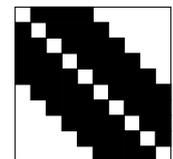
\begin{wrapfigure}{r}{0.265\columnwidth}
\setlength{\columnsep}{-2pt}%
  \vspace{-21pt}
  \begin{center}
    \begin{tikzpicture}[scale=2.05]
    \drawmatr{0}{0}{0}{1}{1}{white}
    \drawmatr{0}{-0.1}{0}{0.1}{0.4}{black}
    \drawmatr{0.1}{-0.2}{0}{0.1}{0.4}{black}
    \drawmatr{0.2}{-0.3}{0}{0.1}{0.4}{black}
    \drawmatr{0.3}{-0.4}{0}{0.1}{0.4}{black}
    \drawmatr{0.4}{-0.5}{0}{0.1}{0.4}{black}
    \drawmatr{0.5}{-0.6}{0}{0.1}{0.4}{black}
    \drawmatr{0.6}{-0.7}{0}{0.1}{0.3}{black}
    \drawmatr{0.7}{-0.8}{0}{0.1}{0.2}{black}
    \drawmatr{0.8}{-0.9}{0}{0.1}{0.1}{black}
    \drawmatr{0.1}{0}{0}{0.4}{0.1}{black}
    \drawmatr{0.2}{-0.1}{0}{0.4}{0.1}{black}
    \drawmatr{0.3}{-0.2}{0}{0.4}{0.1}{black}
    \drawmatr{0.4}{-0.3}{0}{0.4}{0.1}{black}
    \drawmatr{0.5}{-0.4}{0}{0.4}{0.1}{black}
    \drawmatr{0.6}{-0.5}{0}{0.4}{0.1}{black}
    \drawmatr{0.7}{-0.6}{0}{0.3}{0.1}{black}
    \drawmatr{0.8}{-0.7}{0}{0.2}{0.1}{black}
    \drawmatr{0.9}{-0.8}{0}{0.1}{0.1}{black}
\end{tikzpicture}
\end{center}
  \setlength{\abovecaptionskip}{-6pt}%
  \setlength{\belowcaptionskip}{-13pt}%
  \caption{Kernel of size 10}
\end{wrapfigure}
This kernel is a square binary matrix, whose entries are non-zero only on the lower and upper 4 sub-diagonals around the main diagonal (Figure 6). In other terms, denoting $k_{ij}$ the kernel elements, $k_{ij} = 1$ if $1 \leq |i - j| \leq 4$. Otherwise, $k_{ij} = 0$.

For every possible segment $(b_1, b_2)$, a kernel of size $n = b_2 - b_1 + 1$ is convolved with the corresponding autosimilarity sub-matrix (restricted to the bars between $b_1$ and $b_2$) 
which is then normalized by the size of the segment. This leads to a raw convolution score: $c_{b_1,b_2} = \frac{1}{n}\sum_{i,j = 0}^{n - 1}k_{ij}  a_{i + b_1, j + b_1}$. The kernel aims at detecting local 
similarities within the 8 bars surrounding each bar. 
The more similar this surrounding is, the higher the score.

In addition, we combine the kernel score with a regularity penalty $p(n)$, depending on the size $n$ of the segment. Indeed, in pop music in general (and in the MIREX10 RWC Pop annotations in particular~\cite{bimbot2014semiotic}), the distribution of musical segment sizes (in bars) tend to be centered around 8, and they are more likely to be even than odd. In the experiments reported in the next section, we set empirically, $p(8) = 0$, $p(n) = \frac{1}{4}$, if $n$ is a multiple of 4, $p(n) = \frac{1}{2}$ if $n$ is a multiple of 2, and $p(n) = 1$ if $n$ is odd. This penalty modifies the raw convolution score as follows:
\begin{equation}
    c'_{b_1,b_2} = c_{b_1,b_2} - \lambda p(n)  c_{k8}^{max}
\end{equation}
where $c_{k8}^{max}$ is the maximum of the raw convolution score over all restrictions of size 8 bars within the piece, in order to cope with potential discrepancies in sparsity across autosimilarity matrices for different pieces.
In practice, 
$\lambda$ is fitted by cross-validation.

Finally, segment boundaries are found by a dynamic programming algorithm, inspired from \cite{sargent2016estimating}: it keeps the sequence of segments maximizing the global cost defined as the sum of all segment costs.


\section{Experiments}


The proposed method 
was applied to the $\tilde{Q}^T$ representation and tested on the \enquote{structural segmentation} task, as defined in the MIREX campaigns~\cite{downie2008mirex}, on the 
100 songs from the RWC Pop database~\cite{goto2002rwc}. 
MIREX10 annotations~\cite{bimbot2014semiotic} serve as the reference segmentation (1680 segments).
We compare our results with 
state-of-the-art methods listed below.

\subsection{Related Work}

In the context of structural segmentation, numerous methods try to detect segment boundaries from the autosimilarity matrix, or from an \enquote{affinity matrix} derived from it.

The use of autosimilarity for segmenting music structure probably traces back to Foote~\cite{foote2000automatic}. In this work, structural boundaries are detected by applying a kernel along the diagonal, as 
described above. Foote's kernel though aims at detecting ``novelty'' in the signal's autosimilarity matrix, by comparing inter-similarity between the near past and near future at the current point. A high novelty should indicate a low inter-similarity between past and future, hinting towards a boundary between segments.

More recently, convex NMF
 was used for segmenting a pre-processed autosimilarity matrix~\cite{nieto2013convex}. A variant of NMF decomposition is used to enforce the feature space (here, similarity between different bars) to be contracted in convex combinations of columns of the autosimilarity matrix. Factorization results are thus interpreted as the most similar bars, which can then be processed into sections.

Spectral clustering can also be used. 
In~\cite{mcfee2014analyzing}, an affinity matrix is computed from the signal, where the similarity is obtained with k-nearest neighbors and time-proximity rules. Then, interpreting this matrix as a graph, and its values as vertices connectivity, this method studies the eigenvectors of its Laplacian. These eigenvectors can be interpreted as principally connected vertices, forming cluster classes for segmentation.

\begin{table*}[ht!]
 \begin{center}
 \begin{tabular}{|l|l||ccc|ccc|}
  \hline
  \multicolumn{2}{|c||}{Algorithm} & $P_{0.5}$ & $R_{0.5}$ & $F_{0.5}$ & $P_{3}$ & $R_{3}$ & $F_{3}$ \\
  \hline
  \multicolumn{2}{|l||}{NTD-based autosimilarity} & 53.3\% &62.1\%&56.6\%& 66.8\%& 78.1\%&71.1\%\\
  \hline
  \multicolumn{2}{|l||}{Barwise chromagram autosimilarity} & 43.1\% & 45.7\%& 43.9\% & 64.8\% & 68.0\%& 65.8\%\\
  \hline
  Foote & Original &29.7\%&22.3\%&25.1\%& 63.9\%&48.6\%&54.5\%\\
  Novelty~\cite{foote2000automatic} & Aligned on downbeats &42.0\%&30.0\%&34.5\%& 67.1\%&47.7\%&55.0\% \\
 \hline
  \multirow{2}{*}{CNMF~\cite{nieto2013convex}} & Original &22.8\%&21.5\%&21.5\%& 46.8\%&45.1\%&44.7\%\\
  & Aligned on downbeats &31.6\%&28.1\%&28.8\%& 50.7\%&45.4\%&46.5\%\\
  \hline
  Spectral & Original &31.2\%&30.5\%&29.4\%& 60.7\%&60.8\%&58.1\%\\
  Clustering~\cite{mcfee2014analyzing} & Aligned on downbeats &49.2\%&45.0\%&45.0\%& 65.5\%&60.6\%&60.3\%\\
  \hline
 \end{tabular}
\end{center}
 \vspace{-13pt}
 \caption{\small Averaged segmentation scores, and their comparison with several ``blind'' reference methods.}
  \vspace{-4pt}
  \setlength{\belowcaptionskip}{-30pt}%

 \label{tab:seg_scores}
\end{table*}

\begin{table*}[t]
 \begin{center}
 \begin{tabular}{|l||ccc|ccc|}
  \hline
 \multicolumn{1}{|c||}{Algorithm} & $P_{0.5}$ & $R_{0.5}$ & $F_{0.5}$ & $P_{3}$ & $R_{3}$ & $F_{3}$ \\
  \hline
  NTD, with ``oracle ranks'' for each song & 67.1\% & 78.2\%&71.5\%& 78.5\%&90.2\%&83.1\%\\
  \hline
 Neural Networks~\cite{grill2015music}, results from MIREX 2015~\cite{mirex_struc_2015} &80.4\%&62.7\%&69.7\% &91.9\%&71.1\%&79.3\%\\
  \hline
 \end{tabular}
\end{center}
 \vspace{-12pt}
 \caption{\small Averaged segmentation scores in the ``oracle ranks'' condition, compared to the current state-of-the-art (non-blind) method.}
 \label{tab:oracle_scores}
 \vspace{-10pt}
  \setlength{\belowcaptionskip}{-12pt}%

\end{table*}

We primarily compare the NTD method with these techniques for two reasons. First, they are implemented in the MSAF toolbox~\cite{nieto2016systematic}
. Second, they reach state-of-the-art performance among ``blind'' methods for structural segmentation, \textit{i.e.} methods which, like NTD, do not resort to extensive training from examples.  
Note that the segmentation results we obtained with MSAF, though, are slightly worse ($\approx 3/4\%$) than those obtained at MIREX 2016~\cite{mirex_struc_2016}, possibly due to evolutions of the toolbox itself in the interval. We did not tune any of the default parameters.

As current state-of-the-art, we selected the algorithm from~\cite{grill2015music} since it ranked first in this task in the last MIREX campaigns.
However, as opposed to the previous methods, 
it requires supervised training from many examples.

\subsection{Downbeat-Synchronous Alignment}

By construction, the boundaries estimated by the NTD-based approach are aligned on downbeats, which is not the case for the techniques we use as baseline comparisons. As segments generally start and end on downbeats of the song, this alignment could induce a bias favouring our technique. To compensate for this, in addition to the segmentation scores computed with the original boundaries, we compute the scores after having aligned boundaries on the closest downbeat. We call this condition ``Aligned on downbeats''.

In addition, we also processed the barwise autosimilarity obtained directly from the chromagram, in order to measure the impact of the NTD-derived representation vs the raw time-frequency representation.


\subsection{Implementation Details}
RWC Pop signals are sampled at 44100Hz. Bars were estimated by the madmom toolbox~\cite{madmom}. Chromas were extracted from the Constant-Q Transform of the signal with 32-sample hop using Librosa~\cite{mcfee2015librosa}, then mapped to 96 equally spaced chroma vectors per bar. This results in a chromagram $X$ with $12$ rows and $96 \times B$ columns.
Tensors were handled with the Tensorly toolbox~\cite{kossaifi2019tensorly}. We use our own implementation of the NTD algorithm (see Section~\ref{sec:NTD_algorithm}). Segmentation performance was computed with the mir\_eval toolbox~\cite{raffel2014mireval}.

\subsection{Results}
\label{sec:final_results}
Segmentation performance is evaluated with metrics based on \enquote{hit-rate}. The hit-rate considers a boundary as correct if it coincides with a boundary in the reference segmentation within some time window. 
From the count of correct and incorrect segment boundaries, we compute Precision, Recall and F-measure. Tolerance windows were chosen to be 0.5s and 3s, in line with MIREX standards.

As mentioned in Section~\ref{sec:rank_selection}, the ranks of the NTD strongly influence the decomposition and, consequently, the segmentation results. 
Ranks $T'$ and $B'$ are treated as adjustable parameters, and can vary between 12 to 48, with a step of 4. $W$ is fixed to the 12-size identity matrix. The impact of $T'$ and $B'$ is investigated under two rank selection conditions.

In the first condition (Table~\ref{tab:seg_scores}), the RWC Pop data set is divided in two subsets (songs with odd vs even ID number), 
which are alternatively used as tuning (for global optimization of the ranks and the penalty parameter $\lambda$) and test data sets, in a 2-fold cross-validation fashion. Results shown in the table are averaged over the two folds. Hence, in this condition, all songs of a test data subset are decomposed with the same ranks, 
namely $T', B' = 40, 28$ for odd songs, and $48, 24$ for even ones.

In the second condition, presented in Table~\ref{tab:oracle_scores}, the NTD ranks are fitted a posteriori on each song individually: for each tolerance value, separately, we select the ranks leading to the best F-measure for the given song. This is called the ``oracle ranks'' condition, 
corresponding to the situation where a ``perfect'' rank selection procedure would exist. Resulting scores provide an (optimistic) performance upper bound. 

These two tables exhibit very competitive results. In the first (and most realistic) condition, NTD-based segmentation performance exceeds those of the reference ``blind'' methods segmentation. 
In the ``oracle ranks'' condition, the NTD provides higher F-measures than the state-of-the-art, showing strong potential for the technique, provided an efficient rank selection method is eventually developed.

\vfill
\section{Conclusion and Future Work}

Designing relevant audio features from music remains one of the key questions in many MIR tasks. In this paper, we have proposed a three-way tensor representation of music in frequency, short-term (frames) and mid-term (bars), and means to decompose it under the low-rank Nonnegative Tucker Decomposition (NTD) model. This decomposition turns out to be able to provide a compressed representation of interest, capturing salient patterns in music.

We have illustrated the benefits of the method in a structural segmentation task. The NTD-based representation allows to compute a new type of autosimilarity matrix which exhibits a better contrast than those directly computed on 2D time-frequency representations, and seems well-suited to identify musical patterns at the \enquote{right} time-scale for the task. Experimental results are promising and show a potential to compete with state-of-the-art approaches, may they be \enquote{blind}, or greedier on training data.

Additional research is required to consolidate 
the technique. First, as our experiments show, a rank selection criterion would drastically improve segmentation performance. Second, the model does not yet incorporate the notion of proximity between patterns themselves. In parallel, a number of theoretical questions on model identifiability and algorithmic convergence also remain open.


\newpage
\bibliography{ISMIR2020template}

\end{document}